\documentclass[fleqn,usenatbib]{mnras}

\usepackage{newtxtext,newtxmath}

\usepackage[T1]{fontenc}
\usepackage{ae,aecompl}

\usepackage{graphicx}	
\usepackage{amsmath}	
\usepackage{subfig}
\usepackage[font=small,labelfont=bf]{caption} 

\title[]
{Spectroscopic and photometric time series of the bright RRc star T\,Sex
}

\author[Benk\H{o} et al.]{
J\'ozsef M. Benk\H{o}$^{1,2}$\thanks{E-mail: benko@konkoly.hu},
\'Ad\'am S\'odor$^{1,2}$
and Andr\'as P\'al$^{1}$
\\
$^{1}$Konkoly Observatory, Research Centre for Astronomy and Earth Sciences, 
Konkoly Thege M. u. 15-17., H-1121 Budapest, Hungary\\
$^{2}$MTA CSFK Lend\"ulet Near-Field Cosmology Research Group
}

\date{
Accepted 2020 October 30. Received 2020 October 30; in original form 2020 September 24
}

\pubyear{2020}

\begin{document}
\label{firstpage}
\pagerange{\pageref{firstpage}--\pageref{lastpage}}
\maketitle

\begin{abstract}
We present spectroscopic time series observations on one of the brightest northern RRc star, T\,Sex.
Additionally, we also analysed extended photometric data sets, particularly the recent observations of the {\it TESS} space telescope.
The main findings of our studies are as follows: 
T\,Sex, unlike all RRc stars whose space photometry has been analysed, shows 
only the $0.5f_x$ frequency as an additional pulsation frequency. 
With this, T\,Sex may be the first represent of such rare RRc stars found from space photometry.
The spectroscopic data show a periodic distortion of the H$\alpha$ line with the pulsation phase. 
This phenomenon has not been reported for any  RR Lyrae stars.
The characteristic line distortion is probably caused by the turbulent convection, which resulted 
in higher macroturbulent velocity for T\,Sex than for typical RRab stars.
Line doubling of the Na D line was observed between the 0.37 and 0.80 pulsation phases. 
The explanation of this phenomenon is that the two absorption components originate from different sources. 
The redder component comes from the pulsating atmosphere of the star, while the bluer
one from the interstellar space.
At phase 0.438, we detected emission on the Na D line, which may indicate a 
weak shock wave.
\end{abstract}




\section{Introduction}

Variable stars are typically investigated by photometric time series. 
Spectroscopic and, in particular, spectroscopic time series analyses are less frequent. 
It has practical reasons; spectroscopy is generally a more ``expensive genre''.
For a given star, spectroscopy requires a larger telescope, more complex and more expensive equipment
than photometry. This general difficulty is even more serious for RR Lyrae stars because their 
relatively faint apparent magnitudes ($m_v > 7.45$~mag), short periods (0.3\,--\,0.7~d) and 
non-sinusoidal light variations limit the feasible integration time.
The actual situation was well described by \citet{Jurcsik17}:
``Complete radial velocity curves were published for less than 50 Galactic field
RR\,Lyrae stars and less than 10 RR\,Lyraes in globular clusters previously.''
The circumstances, however, are gradually changing as modern echelle spectrographs become more and 
more prevalent. These powerful tools provide useful data on RR\,Lyrae stars even with 
relatively small telescopes.
This is demonstrated nicely by some recent spectroscopic time series studies 
(e.g. \citealt{ChadidPreston17, Sneden17, Gillet19}).

Spectroscopic time series studies of RRc stars -- RR Lyrae pulsating in their first overtone mode -- 
are even less frequent than studies of fundamental-mode pulsator RRab stars,
though RRc stars are by no means less interesting objects.
\citet{Olech09} reported a new class of double pulsating RR\,Lyrae
stars when they discovered two RRc stars in the globular cluster $\omega$~Cen which pulsate with an additional mode beyond their dominant radial first overtone mode.
This surprising new phenomenon has been identified
in all RRc stars by analysing {\it Kepler} and {\it CoRoT} space photometric data \citep{Moskalik13, Moskalik15, Szabo14, Sodor17}.
The period ratio of these additional modes with the dominant (overtone) one $P_x/P_1$ are always in a narrow range around 0.61 or 0.63. 
Soon, further additional modes were
also found in ground-based observations of some RRc stars 
at period ratios of 0.68 \citep{Netzel15} and 0.72 \citep{Prudil17}.
On the basis of his simplified model calculations,
\citet{Dziembowski16} suggested that the additional frequencies of the first two groups 
($P_x/P_1\sim 0.61$, $P_x/P_1\sim 0.63$) might be associated with $l=8$ or $l=9$ non-radial modes. 
The nature of the other two groups are still mysterious.
We have to stress that such kind of additional modes have never been detected in any of 
the RRab stars. Additional modes were also discovered in RRab stars, but at different period
ratios, and those have different explanations. 
The additional modes of RRc stars appear in classical 
double mode (RRd) stars as well. In fact, the first space observation for such modes occurred in the the RRd star AQ\,Leo \citep{Gruberbauer07}. An overtone pulsation appears to be necessary for the excitation of these modes.

These new phenomena have drawn our attention to RRc stars. 
Since space photometric results suggested that each RRc star shows extra modes, the target selection appeared to be an easy task.
According to \citet{Maintz05}, only six RRc stars brighter than ten visual apparent magnitude are known in the northern sky. The brightest one (V764\,Mon, $V_{\mathrm{max}}=7.13$~mag.  $P_1=0.29$~d) is
even brighter than RR\,Lyr itself. Therefore, we focused on this star but we also selected another,
slightly fainter object, T\,Sex ($V_{\mathrm{max}}=9.81$~mag, $P_1=0.3247$~d) as secondary target 
in the observing window of V764\,Mon. We sought to achieve
the most complete pulsation-phase coverage with the shortest possible integration times
for both stars.

A quick look at the spectra revealed that V764 Mon is not, in fact, an RR\,Lyae, 
but a fast-rotating $\delta$ Scuti star. The results about this star will be published elsewhere. 
In this paper, we study T\,Sex, a bona fide RRc pulsator.

\section{Observations and reduction}

\subsection{Photometric data}

Two sufficiently extensive photometric time series were analysed.
The ASAS-3 {\it V} band data (All Sky Automated Survey, \citealt{ASAS}) contains 504 observed data points.
Before the analysis, the outlying points fainter than 10.4~mag 
and the less accurate observations flagged by `D' in the data base had been removed. In the end, 457 data points remained.

Up to the time of writing of this manuscript, the {\it TESS} space mission \citep{Ricker15} 
observed T\,Sex once, in Sector~8, obtaining 17\,755 data points.
The observation of Sector~8 was taken in February 2019 almost continuously, covering 24.62 days with 2~min exposures. 
This exposure time represents oversampled high-cadence
observations\footnote{The data are publicly available at
the Mikulski Archive for Space Telescopes: {\url{https://mast.stsci.edu/portal/Mashup/Clients/Mast/Portal.html}} 
}.

From the data offered by the archive, the light curves obtained from the corrected aperture 
photometry (PDCSAP) fluxes were used for this analysis, and only the best-quality data 
(marked with a quality flag 0) were used, which corresponds to 13\,395 data points. The fluxes were transformed to a magnitude scale.
We mention that {\it TESS} magnitudes (zero point, the amplitude and the errors) 
are scaled with the reference magnitude. 
We accepted the value of $m_{\mathrm{TESS}}=9.779$~mag according to the TESS data release.
The typical error of the individual photometric data points is $\sim0.0013$~mag.

\begin{table}
        \centering
        \caption{
Log of T\,Sex  spectroscopic observations
}
        \label{tab:log}
        \begin{tabular}{ccccr} 
                \hline
  Night &        JD             & $\langle S/N \rangle$ &  $\phi$ & $n$\\
     (yyyy-mm-dd) &  (-2 400 000)  &       &         &     \\
                \hline
 2015-03-05   &   57087        & 11   &     0.606-0.732     & 2  \\
 2015-03-07   &   57089        & 51   &     0.003-0.932     & 14  \\
2015-03-10   &   57092        & 19    &     0.020-0.945     & 7  \\
                \hline
        \end{tabular}
\end{table}

\subsection{Spectroscopic data}

For the observations we used the ACE fibre-fed \'echelle spectrograph
attached to the 1-m RCC telescope at the Piszk\'estet\H{o} mountain station of the Konkoly Observatory.
The spectra cover the 4150-9150~\text{\AA} wavelength range with a resolution of $R\approx$20\,000.
A total of 23 spectra were recorded from the target star on three nights between 
5 and 10 March 2015 (Table~\ref{tab:log}). 
The integration time was 30~min, which is a good trade off between 
getting enough pulsation phase resolution and reaching acceptable signal-to-noise ratio 
(see column 4 in Table~\ref{tab:log} for nightly averaged $S/N$ values estimated by {\sc{iSpec}}
\citep{Blanco-Cuaresma14, Blanco-Cuaresma19}). 

The ACE spectra were reduced using standard {\sc IRAF} \citep{iraf1,iraf2} tasks including bias, 
dark and flat-field corrections, aperture extraction, and wavelength calibration 
using thorium-argon calibration images, taken after every third object frames.
The normalisation,  cosmic-ray  filtering,  order  merging were performed by our
 {\sc python} scripts (developed  by \'AS).  Each spectrum was also corrected to the barycentric frame.

\section{Analysis of the photometric data}\label{sec:phot}

Although this work focuses on spectroscopic time series observations of T\,Sex,
we also needed some photometric data for determining the proper period and phases.
Furthermore, the analysis of the photometric data yielded an unexpected result, too.

\begin{figure}
\includegraphics[angle=0,scale=.47, trim=0 0 0 0]{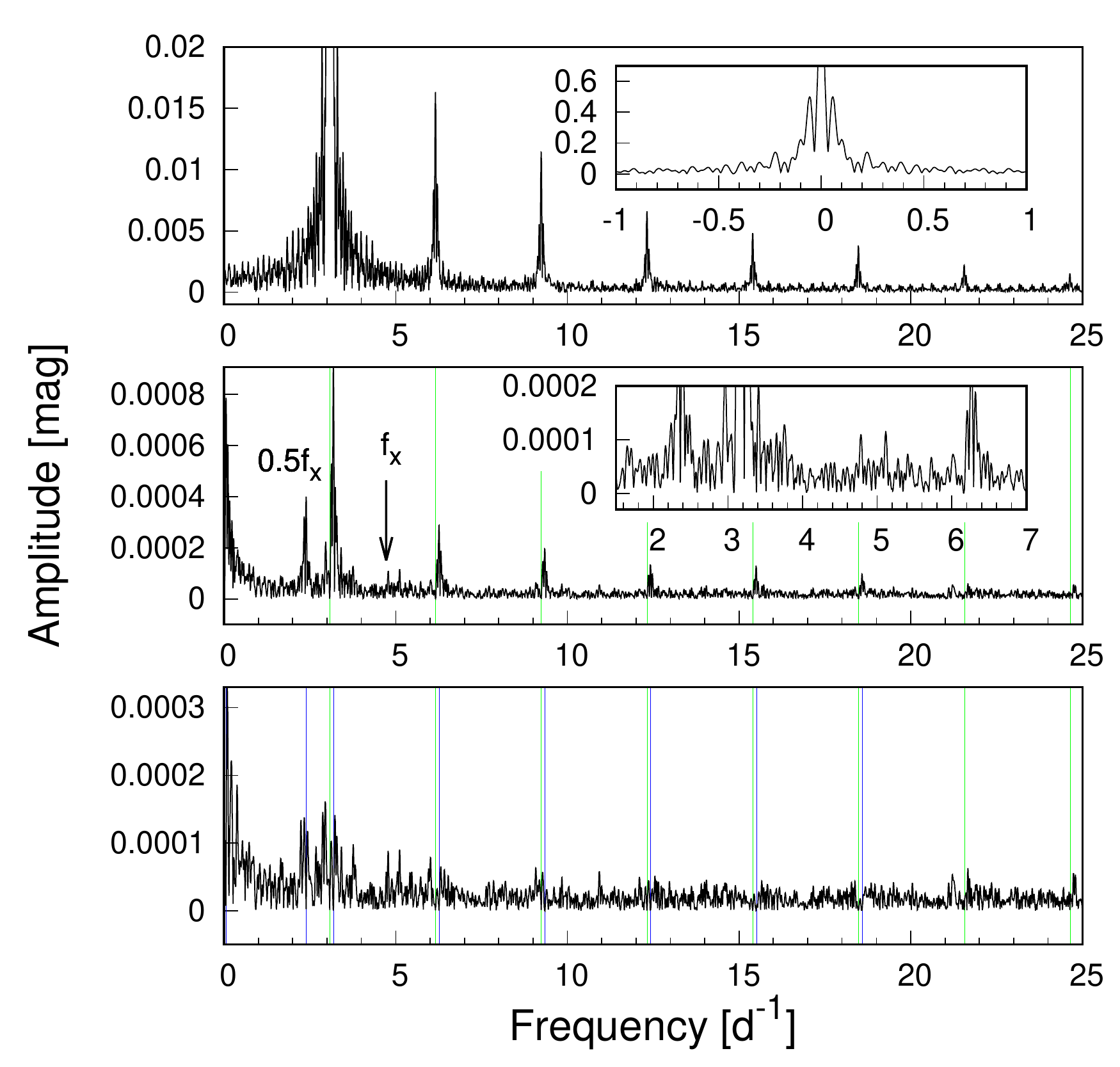}
\caption{Fourier spectra of the {\it TESS} data on T\,Sex during the subsequent pre-whitening steps. Top panel: 
The original spectrum
and the central part of the window function (inset). Middle panel: the spectrum after pre-whitening the data with the
main pulsation frequency $f_1$ and its 16 harmonics.
Thin green vertical lines show
the position of the pre-whitened frequencies.
The inset in the middle panel is the expansion of the frequency range around $0.5f_x$ and $f_x$. Bottom panel: 
residual after pre-whitening the spectrum shown in the middle panel with eight further significant 
frequencies (indicated with blue vertical lines). To make the harmonic structure more apparent, the amplitude limit on the top panel 
is lower than the amplitude of the main frequency (0.08~mag).
}\label{fig:phot}
\end{figure}

The ASAS data set was analysed by the Fourier fitting tool 
of the {\sc Period04} package \citep{Period04}. 
Since these data points spread over 9 years (3276 days between December 2000 and November 2009),
the Nyquist frequency ($\sim$0.24~d$^{-1}$)
is significantly lower than the pulsation frequency ($f_1$=3.0798~d$^{-1}$), and
the Fourier spectrum had to be calculated well above this limit frequency. 
Consequently, the resulting spectrum has a periodic structure. It contains the main frequency, 
its daily alias frequencies and the annual aliases caused by the seasonality of the observations
and the daily repetitions.
The 1-year frequency also occurs in side peaks around the main frequency. 
Apart from the pulsation frequency, no additional frequency can be detected above the noise level ($\sim 0.04$~mag).

The Fourier spectrum of the {\it{TESS}} data shows the main pulsation frequency ($f_1$=3.079~d$^{-1}$) 
and its harmonics (see top panel in Fig.~\ref{fig:phot}). 
We pre-whitened the data with $f_1$ and 
its all 16 significant harmonics up to the frequency  50.0~d$^{-1}$.
The residual spectrum is shown in the middle panel of Fig.~\ref{fig:phot}.
Eight significant frequencies can be detected in this residual.
Six of them can be written in the form $kf_1+f^{(1)}$, where $k=1,2,\dots 6$, and $f^{(1)}=0.103$~d$^{-1}$, moreover
$f^{(2)}=0.055$ and $f^{(3)}=2.3919$~d$^{-1}$.
All but one of these frequencies are of instrumental origin.
The window function (inset in top panel of Fig.~\ref{fig:phot}) contains the data length frequency ($f^{(2)}=0.055$~d$^{-1}$). 
We can now identify $f^{(1)}\approx 2f^{(2)}$.
These identifications are within the frequency errors because the 
short {\it TESS} data set the Rayleigh frequency resolution is rather low (0.04~d$^{-1}$).

\begin{figure}
\includegraphics[angle=-90,scale=.33, trim=0 0 2cm 0]{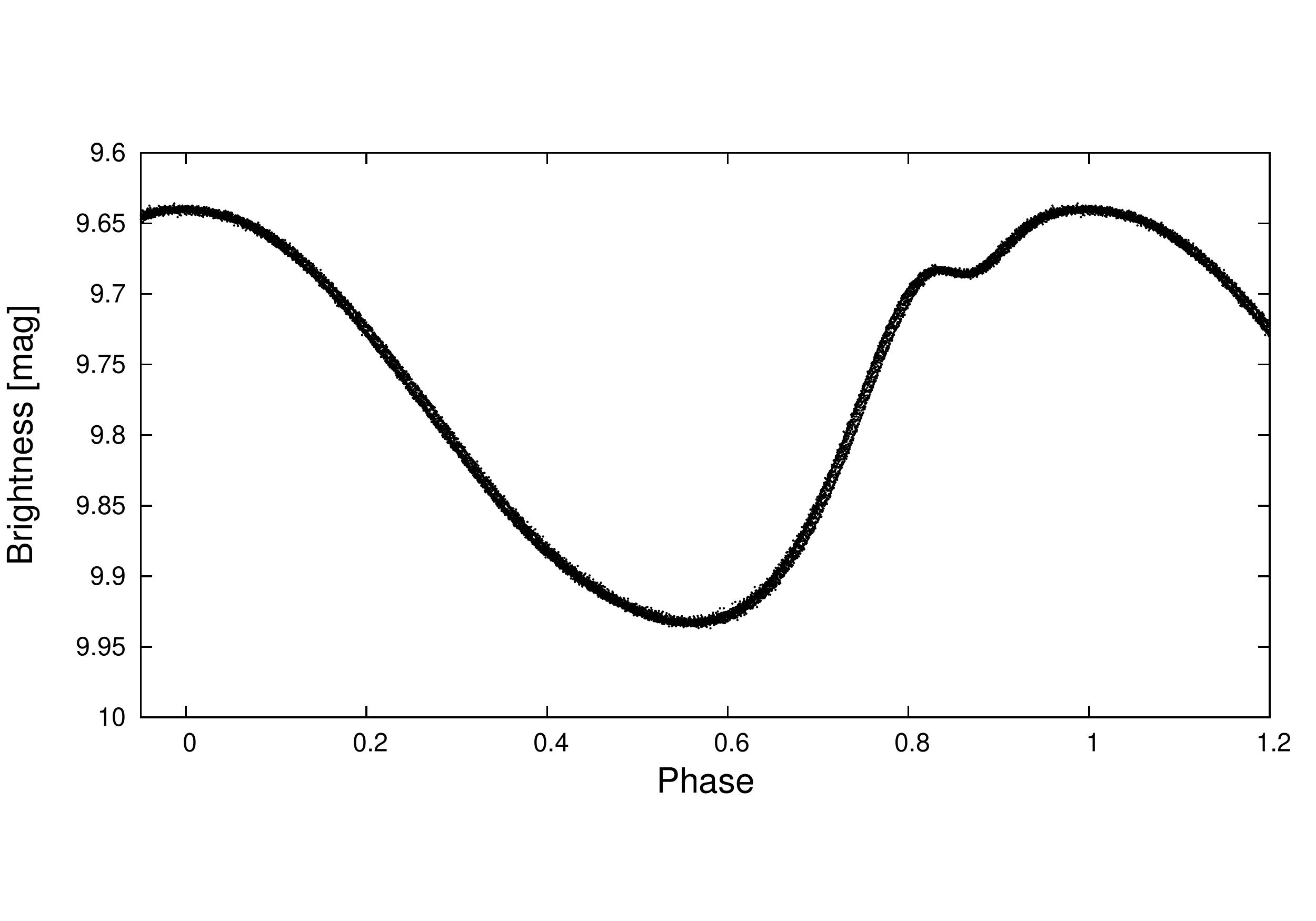}
\caption{
Phased light curve of T\,Sex observed by {\it TESS} satellite.
}\label{fig:phase}
\end{figure}

\begin{figure}
\includegraphics[angle=0,scale=.47, trim=0 0 0 0]{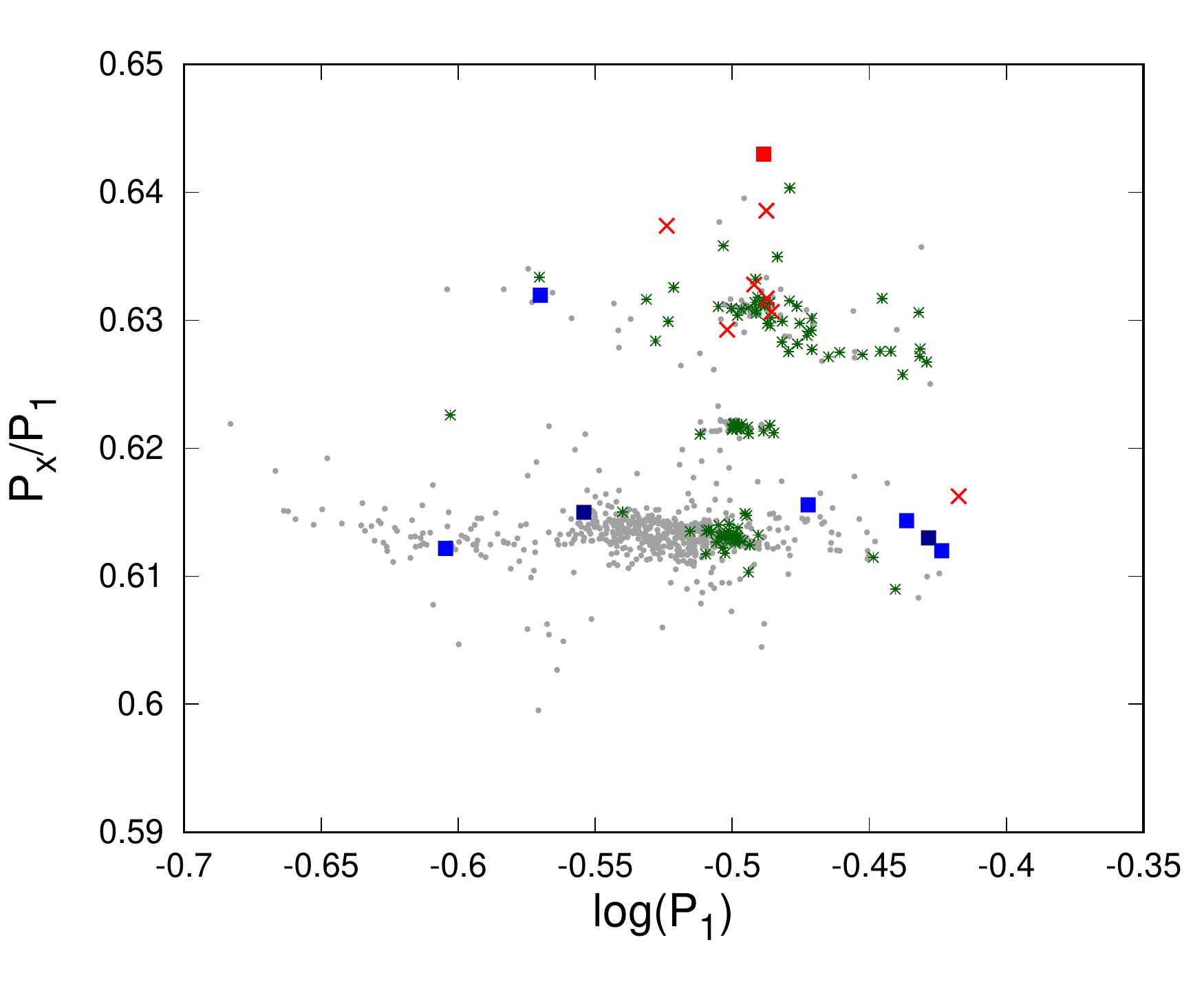}
\caption{Petersen-type diagram of RRc stars showing the additional ($f_x \sim 0.61$) frequencies. The OGLE
RRc stars analysed by \citet{Netzel19}:
grey dots are the RR$_{\mathrm{0.61}}$ sub-sample;
green asterisks denote stars showing $0.5f_x$ frequency as well, while red `x' symbols show stars 
with $0.5f_x$ without $f_x$. Filled squares indicate space results: 
light blue -- \textit{Kepler} \citep{Moskalik15, Sodor17}, dark blue -- \textit{CoRoT} \citep{Szabo14}, red square -- T\,Sex \textit{TESS} (this work).
}\label{fig:petersen}
\end{figure}

The only significant ($S/N=5.7$) non-technical frequency is $f^{(3)}$.
As mentioned in the introduction, all the 
space photometric measurements of RRc stars that have been studied show additional frequencies. 
The Fourier spectra of those stars typically contain a strong peak ($f_x$) with a ratio to the main period $f_1$ of around 0.61 or 0.63 (see blue squares in Fig.~\ref{fig:petersen}), as well as its harmonics and linear combinations with $f_1$, respectively. In some cases, the sub-harmonic $0.5f_x$ is also detectable ({\it CoRoT} \citealt{Szabo14}, {\it Kepler/K2} \citealt{Moskalik15, Molnar15, Sodor17}). 

One possible explanation for the missing $f_x$ frequency of T\,Sex
is that, as the above cited works have shown, the amplitude of $f_x$
can change strongly over time. Perhaps \textit{TESS} measured
the star in a `low amplitude state’, when the amplitude of the frequency
$f_x$ was below the detection limit. We expect \textit{TESS} to
re-observe T\,Sex in February 2021 (Sector 35), and the question
may be decided.
However, \citet{Jurcsik15}, studying RRc
stars of M3 by ground-based multicolour photometry, found that stars
showing the $f_x$ frequency are bluer than those not showing such a signal.
This finding suggests intrinsic physical difference between these two groups 
of RRc stars.

According to \citet{Dziembowski16}, signals at sub-harmonic
frequencies $0.5f_x$, are the real frequencies of the non-radial 
modes of degrees $l=8$ and $l=9$, and the signals at $f_x$ are harmonics. 
Because of cancellation effects, the harmonic generally has better visibility 
than the mode frequency itself. 
On the large OGLE RRc sample, \citet{Netzel19} showed recently that 
the longer-ratio sequence ($P_x/P_1\sim 0.63$) belongs most probably to the $l=8$ mode.
Stars pulsating in this mode tend to show both $f_x$ and $0.5f_x$ frequencies as well.
From the 960 stars in which an additional mode was found by \citet{Netzel19} (grey dots in Fig.~\ref{fig:petersen}), 
$0.5f_x$ was also detected for 114 stars (green asterisks in Fig.~\ref{fig:petersen}) and even in 35 cases (3.6\%), 
this frequency had larger amplitude than that of $f_x$.
For another seven stars (red `x'-es in Fig.~\ref{fig:petersen}), only $0.5f_x$ is detected.

If we identify $f^{(3)}$ as $0.5f_x$ then 
$f_x=4.7826$~d$^{-1}$ and $f_1/f_x=0.643$ (see red filled square in Fig.~\ref{fig:petersen}). This ratio
is higher than the median of the 0.63 sequence but
within the observed range of this ratio.
Such identification of $f^{(3)}$ is likely because a peak, although not significant ($S/N=3.2$), is indeed visible at the position of the calculated $f_x$ (see the inset in middle panel of Fig~\ref{fig:phot}).
Near to this one, at $f^{(4)}=5.1132$~d$^{-1}$, we can also see a peak of similar amplitude ($S/N=3.1$). The ratio of this 
frequency ($f_1/f^{(4)}=0.602$) suggests that it might be the harmonic of the $l=9$ mode. That is, T\,Sex contains
both $l=8$ and $l=9$ mode pulsations as well. This is
rather common phenomenon, \citet{Netzel19} found this 
in more than 10\% of their sample.

By removing the eight significant frequencies discussed above with a subsequent pre-whitening step, 
the residual spectrum shown in the bottom panel of Fig.~\ref{fig:phot} is obtained.
No further significant ($S/N > 4$) frequency can be detected in this spectrum
but some low frequency excess can be seen. It has at least two sources: (i) a global instrumental trend and (ii)
a cycle-to-cycle light curve variation similar to the found for RRab stars
(see in Sec.4.1 in \citealt{Benko19} for a discussion) since both phenomena are clearly visible in the \textit{TESS} light curve of T\,Sex.

The pulsation period obtained from the {\it TESS} data ($P_1=0.3248\pm0.004$~d) is in agreement with the 
more precisely defined ASAS period ($P_1=0.324696\pm0.00003$~d).
The {\it TESS} light curve folded with the ASAS period is shown in Fig.~\ref{fig:phase}.
Since the standard deviation of the curve is very small, no phase-shifted cycles are seen, it is likely that the period between the end of the ASAS measurements (2009) and the beginning of the TESS measurements (2019) did not change significantly.

\section{Radial velocity curve}\label{sec:vrad}

\begin{table}
        \centering
        \caption{Sample of the radial velocity data tables. The columns contain the 
barycentric Julian date (BJD), the measured radial velocity $v_{\mathrm{rad}}$, its uncertainty $\sigma (v_{\mathrm{rad}})$
and the corresponding pulsation phase $\phi$.  
}
        \label{tab:radseb}
        \begin{tabular}{ccccc} 
                \hline
 BJD &   $v_{\mathrm{rad}}$  &  $\sigma (v_{\mathrm{rad}})$ & phase\\
 (d) &    (km s$^{-1}$)      &  (km s$^{-1}$) & \\
                \hline
2457089.25160  & 22.87   &  1.67&  0.142584    \\
2457089.27272  & 28.37   &  1.95&  0.207629  \\
2457089.29368  & 31.76   &  3.38&  0.272182   \\
2457089.32674  & 35.25   &  1.60&  0.374000   \\
2457089.34768  & 37.48   &  2.31&  0.438491   \\
(...)  & & & \\
                \hline
        \end{tabular}\label{Tab:vrad}
\end{table}

\begin{figure}
\includegraphics[angle=-90,scale=.33, trim=0 0 0 0]{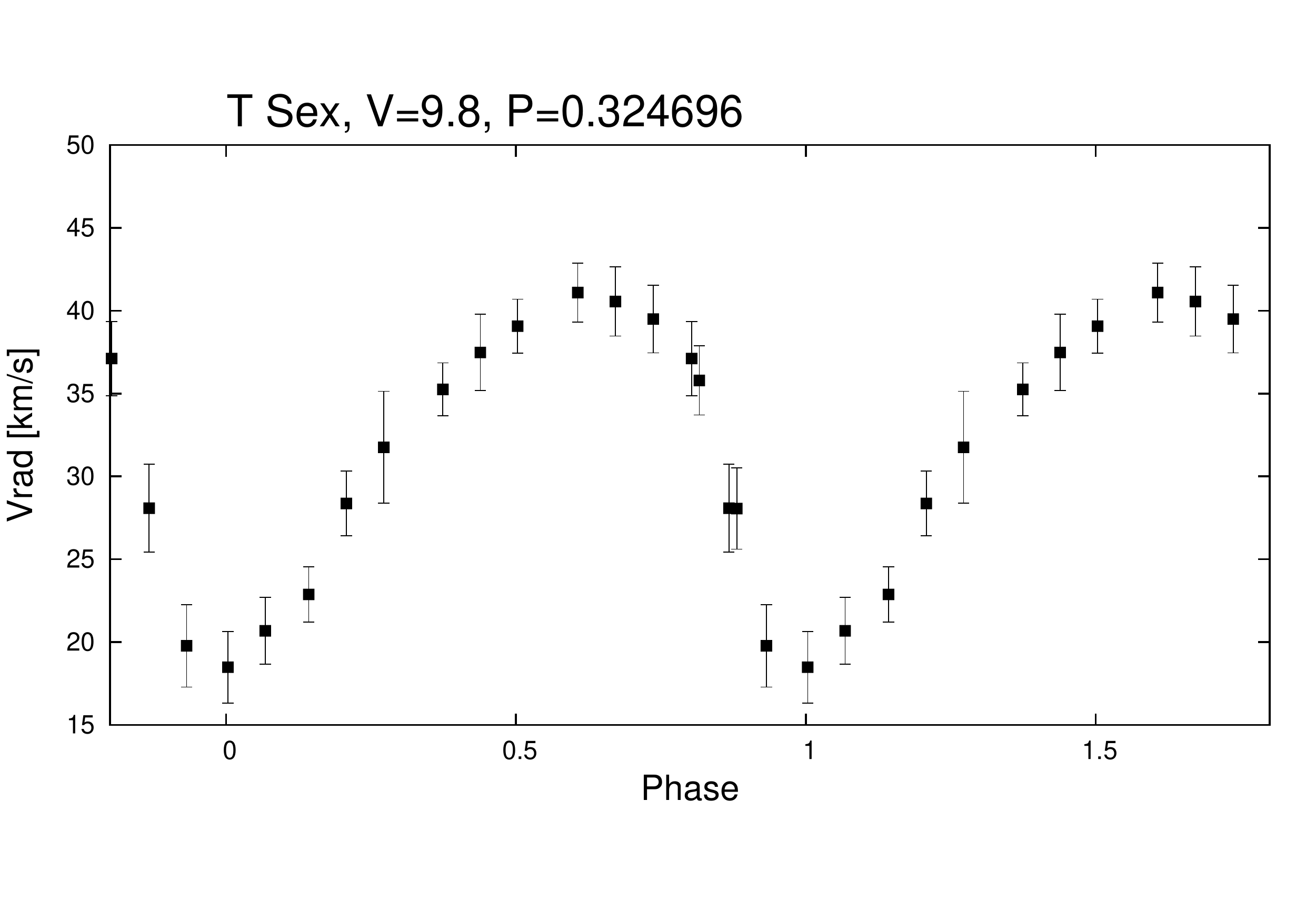}
\caption{Radial velocity curve of T\,Sex.
The error bars show the 1{$\sigma$} calculated formal errors (around 1-2 km~s$^{-1}$). 
}\label{fig:vrad}
\end{figure}

We determined the radial velocity curve of T\,Sex.
The radial velocities were calculated by cross-correlating the spectra with
a metallic line mask containing 622 metallic lines between 4800 and 5600~\text{\AA}.
We used Gaussian fitting functions for determining the radial velocities and their uncertainties from the
the cross-correlation functions. 

The S/N ratio in the cross-correlation functions are 80\,--\,150, which provide a precision 
of $\sim$1\,--\,2~km~s$^{-1}$ for the radial velocities. 
The  systematic  errors resulted in the data processing and the stability of the wavelength 
calibration system of  the ACE instrument  are  better than 
0.36 km s$^{-1}$, based on observations of radial velocity standards \citep{Derekas17}.   
The radial velocity curves are published as electronic tables. The structure of
these tables is shown in the excerpt in Table~\ref{Tab:vrad}.

We used here the ASAS-3 period determined above in Sec.~\ref{sec:phot},
For zero phase of the radial velocity curve we simple used the 
current ephemeris  of GEOS 
database\footnote{\url{http://rr-lyr.irap.omp.eu/dbrr/}} 
\citep{LeBorgne07}
belonging to the photometric maxima because these two values are coincident
within the observation error \citep{Jurcsik15, Jurcsik17, Sneden17}.  
The exact ephemeris was not critical in this study,
but for the sake of completeness, we give it as we
found from our polinomial fit:
$\phi_{v_{\mathrm {rad}}}(0)=2457089.529\pm 0.08$.
The phase difference between radial velocity minimum and
photometric brightness maximum in $V$ is $0.025\pm 0.08$ in 
agreement with the previous study \citep{Sneden17}.
The obtained radial velocity phase curve in Fig.~\ref{fig:vrad}
demonstrates well the complete phase coverage of our observations.

Although T\,Sex is one of the brightest RRc stars,
up to now only five radial velocity curves were published
\citep{Tift58,Preston64,Barnes88, Liu89, Sneden17} and phase coverage of these curves
are complete only in \citet{Tift58}, \citet{Barnes88} and \citet{Liu89}. 
By using the zero point of a four-element 
Fourier fit to the radial velocity curve we determined the mean velocity as 
$v_0=31.7\pm0.5$~km~s$^{-1}$. This mean velocity is approximated
the velocity of the stellar rest frame with respect to the solar system barycenter 
($v_{\gamma}$). Strictly speaking, however, this is not completely true, because the optical depth
changes during the pulsation. The radial-velocity curve does not represent any
physically moving fluid element (e.g. \citealt{Karp75}). However, 
the so-called k-term -- the difference between $v_{\gamma}$ and $v_0$ -- must be less 
than 2~km~s$^{-1}$ for RRc stars. This value was found for more extended atmosphere 
of Cepheids by \citet{Nardetto08}.
 
Our mean velocity value is 6.5~km~s$^{-1}$ higher than the 
latest published in the literature ($v_0=25.2\pm1$~km~s$^{-1}$, \citealt{Gontcharov06}).
This latter mean radial velocity compilation, however, is prepared for
Galactic kinematic purposes and optimised for non-variable stars. As \citet{Kollmeier13}
showed, the mean radial velocity of RRc stars can be well estimated by measuring the 
radial velocity curve at the phase of $\phi=0.32$. By a simple interpolation we obtain
$v(0.32)=33.5\pm2.4$~km~s$^{-1}$. This value is consistent within 1$\sigma$ 
with our previous calculation
and the recent measurements of \citealt{Sneden17}.
(They did not calculated $v_{\gamma}$ for T\,Sex 
because of their incomplete phase coverage, but from their data we found
$v(0.32)=28.6\pm2.5$~km~s$^{-1}$.)

\section{Spectral variations with pulsation phases}

\subsection{Hydrogen H$\alpha$ line}

\begin{figure}%
    \centering
    \includegraphics[width=8.9cm,trim=3cm 0cm 3cm 0]{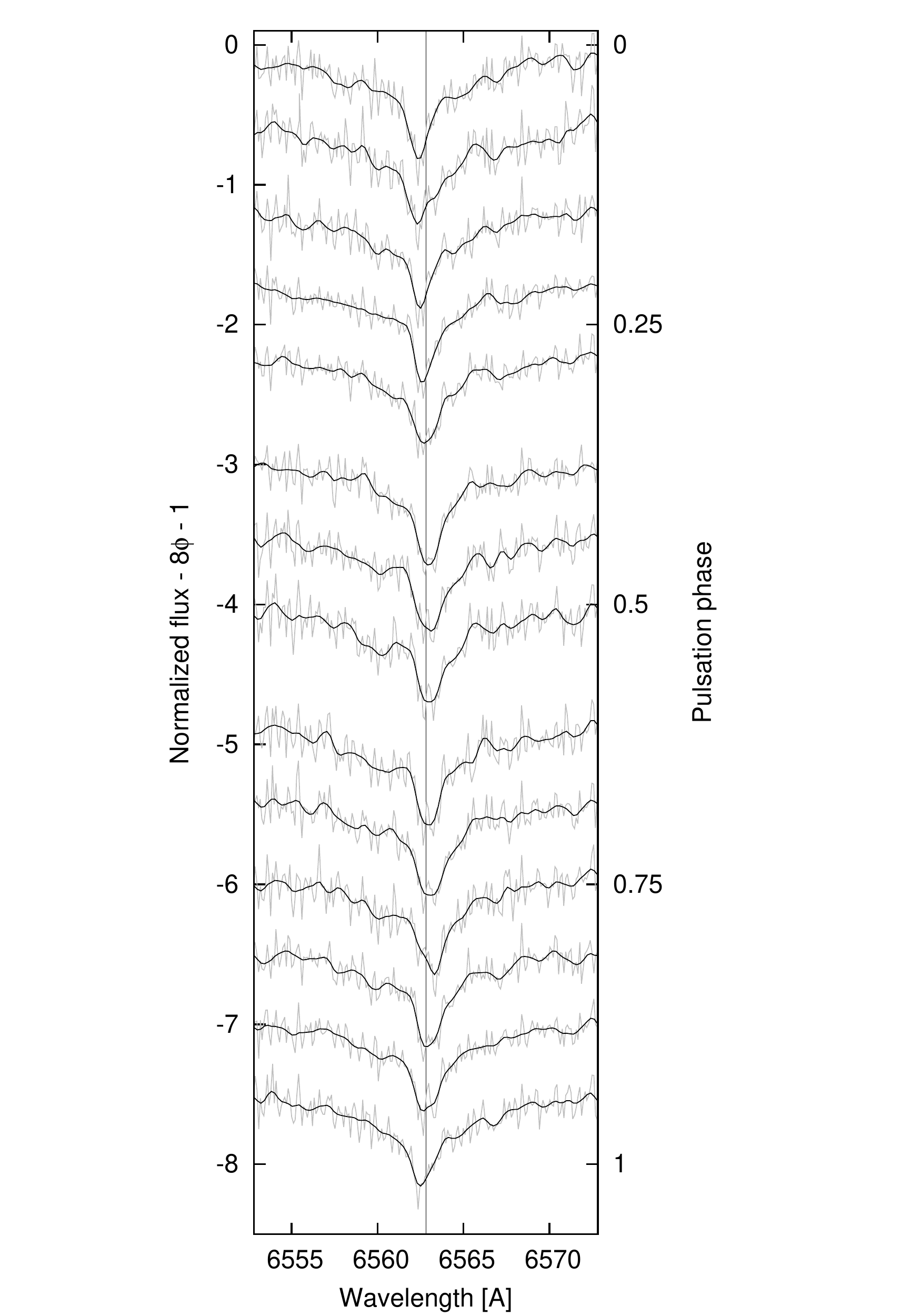}%
\caption{H$\alpha$ line variations of T\,Sex over the pulsation cycle in the stellar rest frame.
The normalised spectra are shown with thin grey line. The black curves are the smoothed spectra. 
Vertical line marks the laboratory position of the H$\alpha$ line.
}\label{fig:Tsex_Ha}%
\end{figure}

Our spectra cover the hydrogen Balmer H$\alpha$, H$\beta$  and H$\gamma$ lines.
Since the detector is more sensitive in redder wavelengths, H$\alpha$ lines have the highest 
S/N ratio among the Balmer series. Therefore, we investigated the motions and line-profile variations 
of the H$\alpha$ line over the pulsation phase. 

In Fig.~\ref{fig:Tsex_Ha}, the H$\alpha$ line-profile variations 
of  T\,Sex on the best night (2015-03-07) are plotted in the stellar rest frame. 
To make the phase dependence easier to follow,
the normalised spectra are shifted vertically with phase dependent constants. The thin grey curves represent
the original spectra. Since these spectra show many weak (mostly telluric) lines and some noise, 
we plot smoothed spectra as well (black curves in Fig.~\ref{fig:Tsex_Ha}). 
The latter are better for following line profile variations.   

If we look at the line positions, we find periodic shifts relative to the stellar rest frame: 
The H$\alpha$ line is blueshifted between
phases $\phi\sim0.87 - 0.21$ and redshifted between  $\phi\sim0.37 - 0.74$. 
This variation is the natural consequence of the radial pulsational motion
in which the hydrogen-absorbing layer is involved and defines a radial velocity
curve slightly different from the one obtained from metallic lines. These differences
are discussed in detail by \citet{Sneden17}. 

\subsection{Periodic line-profile distortions}

The H$\alpha$ line, however, does not simply shift periodically around the laboratory wavelength corrected with the center-of-mass velocity,
but its profile also changes with the pulsation phase.
In Fig.~\ref{fig:Tsex_profile}, we show two highly asymmetric phases compared
with a symmetric one. Such line-profile variations has not been reported for RRc stars before. 

This line asymmetry
is a remarkable difference compared to the variation of the H$\alpha$ line of RR\,Lyrae. In that case, an H$\alpha$ line doubling can be detected between $\phi=0.943$ and
$\phi=1.027$ but the core of the line remains symmetric in all phases \citep{Gillet19}.
Additionally, some metallic lines of  
RR Lyrae also show phase-dependent profile distortions around the
brightness maxima ($\phi\sim 0.91-0.97$) 
and the strength of this effect depends on the Blazhko phase
\citep{Chadid96,Chadid97}. These two phenomena have been explained by the
same physical mechanism: a hydrodinamical shock wave passing through the pulsating atmosphere causing two distinct absorbent layers \citep{Schwarzschild52,Fokin97}. 
In this case, the asymmetric metallic lines would be nothing more than overlapping 
double lines.

The H$\alpha$ line doubling and the asymmetry of metallic lines in RRab stars appear
only around the phase of maximum brightness. 
The phenomenon presented here is more similar to that observed in some metallic lines 
of classical Cepheids and $\beta$~Cep stars \citep{Nardetto08, Nardetto13}.
\citet{Nardetto08} discussed three explanations for the effect.
The time- and wavelength dependence of the limb-darkening within the spectral lines,
velocity gradients in the atmosphere and the relative motion of the line-forming 
region with respect to the corresponding mass elements. 
The third explanation is closely related to the one that was mentioned for RRabs.
\citet{Nardetto08} concluded that for quantitative modelling of line
asymmetries, a detailed hydrodynamic model, in which convection is taken into account, 
should be combined with a wavelength-dependent radiative code.

\subsubsection{Phenomenologic explanation}

Pulsating atmospheric models
that take into account both convection and shock waves and are able
to compute synthetic lines profiles for Cepheids or RR\,Lyrae stars 
have not yet been developed. There is, however, a smart tool, called {\sc NRP Animation Creator (NRPAC)}\footnote{\url{http://staff.not.iac.es/~jht/science/nrpform/}}
\citep{Schrijvers97, Telting97, Schrijvers99} that allows us to model line profile
variations of pulsating stars.
The program is optimised for fast rotating non-radially oscillating variables,
such as $\delta$ Scutis and related stars. Because of its assumptions (e.g. of adiabatic pulsation)
it is not suitable for quantitative analysis of line-profile variations in an RR Lyr pulsator.
Nonetheless, if we assume a radially pulsating ($l=m=0$) star with the parameters of a typical
RRc  star (mass 
$M=0.65$~$M_{\sun}$, 
radius $R=4$~$R_{\sun}$, 
3D velocity amplitude $A(v)$=20~km~s$^{-1}$,
pulsation frequency $f_1$=0.3~d$^{-1}$, 
$v\sin i=15$~km~s$^{-1}$), 
we can qualitatively 
reproduce the observed line-profile distortions.

\begin{figure}%
    \centering
    \includegraphics[width=6.cm,angle=270,trim=0cm 0cm 0cm 0]{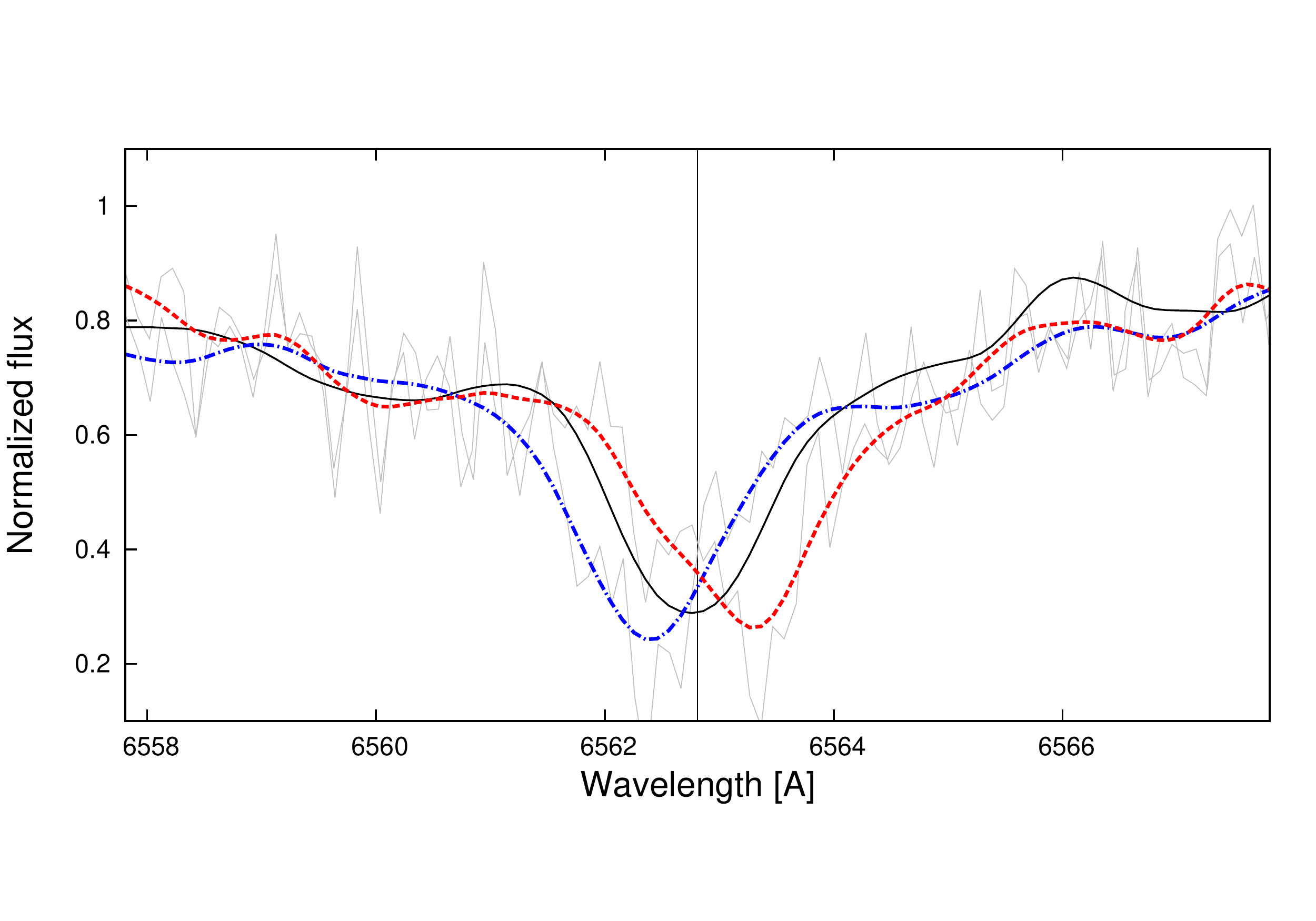}%
\caption{
H$\alpha$ line-profile distortions of T\,Sex over the pulsation cycle.
The normalised spectra are shown with thin grey lines. 
The black continuous curve shows a smoothed spectrum at $\phi=0.607$ with symmetric 
line profile
corrected with its relative velocity (9.39~km~s$^{-1}$) to the stellar rest frame.
The blue dash-dotted curve is a blue shifted asymmetric spectrum at $\phi=0.003$ while the 
red dashed curve shows the red shifted asymmetric spectrum at $\phi=0.737$.   
The vertical line indicates the laboratory position of the H$\alpha$ line.
}\label{fig:Tsex_profile}%
\end{figure}

\begin{figure}
\centering
\subfloat[]{\includegraphics[width=2.2cm,trim=0cm 0cm 0cm 0]{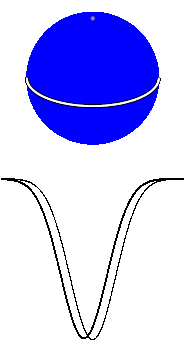}}\quad
\subfloat[]{\includegraphics[width=2.2cm,trim=0cm 0cm 0cm 0]{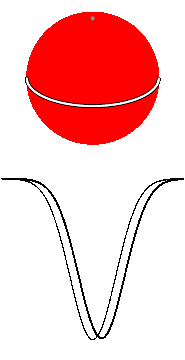}}\quad\\
\subfloat[]{\includegraphics[width=2.2cm,trim=0cm 0cm 0cm 0]{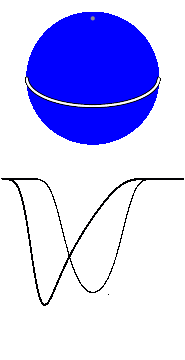}}\quad
\subfloat[]{\includegraphics[width=2.2cm,trim=0cm 0cm 0cm 0]{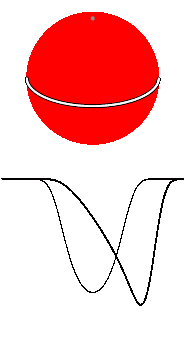}}\quad
\caption{Extreme cases of line profile variation on a radially pulsating star
with the parameters of a typical RRc stars. The figures were prepared by using 
NRP Animation Creator. The subfigures
a) and c) show maximal contraction b) and d) maximal expansion phases. In the top
row $v\sin i=1$~km~s$^{-1}$, in the bottom row $v\sin i=15$~km~s$^{-1}$.}
\label{fig:anim}
\end{figure}   

Fig.~\ref{fig:anim} shows the line profile of two extreme 
phases (the maximal contraction in the left side, and the maximal expansion in the right)
of a radially pulsating star. The line profiles of the resting stars are shown by thin lines.
The same parameters were used for all the plots, except for $v\sin i$.
We used $v\sin{i}=1$~km~s$^{-1}$ for the top row and 15~km~s$^{-1}$ for the bottom row. 
For $v\sin{i}=1$~km~s$^{-1}$, the line profile
is practically not distorted and it shows similar profiles as the observed one of 
\citet{Gillet19} for RR\,Lyrae.  
Fig.~\ref{fig:anim} illustrates well that the key of qualitatively reproducing the observed line-profile variations of T\,Sex is applying a relatively high $v\sin{i}$
value in {\sc{NRPAC}}. 

\citet{Preston19} summarises the present
knowledge of RR\,Lyrae axial rotation and macroturbulence, which are 
generally hard to separate because these two phenomena broaden the line profiles in a similar fashion.
\citet{Preston19} found an upper limit for the macroturbulent velocity of RRab stars 
as $5\pm1$~km~s$^{-1}$. They also showed that this velocity is less uniform for 
RRc stars and it could be as high as $\sim12$~km~s$^{-1}$.

As we have seen, reproducing the line-profile variations of T\,Sex with {\sc{NRPAC}} requires a sufficiently large $v\sin i$. 
This tool, however, does not use the macroturbulent velocity as a free parameter, therefore
the suggested high $v\sin i$ does not necessary mean a high equatorial 
rotation speed, but could imply a more intense macroturbulence caused 
by the convection. This finding agrees with the theoretical calculations of  \citet{Gautschy19},
showing the more significant 
role of atmospheric convection in RRc stars than in RRab stars. 
It means, on the one hand, that convection is more important in the
energy transport and, on the other hand, that, unlike RRab stars, 
it is present in almost all pulsation phases.

The question may arise whether the non-radial mode discussed in
Sec.~\ref{sec:phot} could cause the observed line profile changes or not. Most probably not. First, the expected period of an $l=8$ non-radial is $P_3=1/f^{(3)}=0.41808$~d significantly different from the pulsation period, which is the period of the observed effect.
Second, non-radial modes primarily modify the shape of line cores, while only slightly shift the position of the lines
(see e.g. \citealt{Telting03} and references therein).
However, we see strong shifts in the line position:
there is a difference of $\sim 1$~\text{\AA} between the two extreme positions of the H$\alpha$ line in Fig~\ref{fig:Tsex_profile}, which is typical for a radial mode.
Thirdly, the degree of line distortion expected from an $l=8$ mode is much smaller than that results from the large macroturbulence discussed above. 
If we prepare simulated line profiles by NRCAP assuming a combination of a radial ($l=0$) and an $l=8$ non-radial modes,
the obtained simulated line profiles are indistinguishable from those shown in Fig.~\ref{fig:anim}. 
If we could subtract the variations of the radial mode, and examine only the effects of non-radial mode, our simulated line profiles show that even then 
we obtain a measurable (some percents of) variation only under some
special circumstances, when the inclination is high and the absolute value of the sectorial number $\vert m\vert$ is also high.
In summary, the shape, strength, and period of the line profile variations found do not support the explanation that this would be caused by a non-radial mode.

\subsubsection{Spectral fitting}

\begin{figure}%
    \centering
    \includegraphics[width=3.5cm,angle=270,trim=0cm 0cm 0cm 0cm]{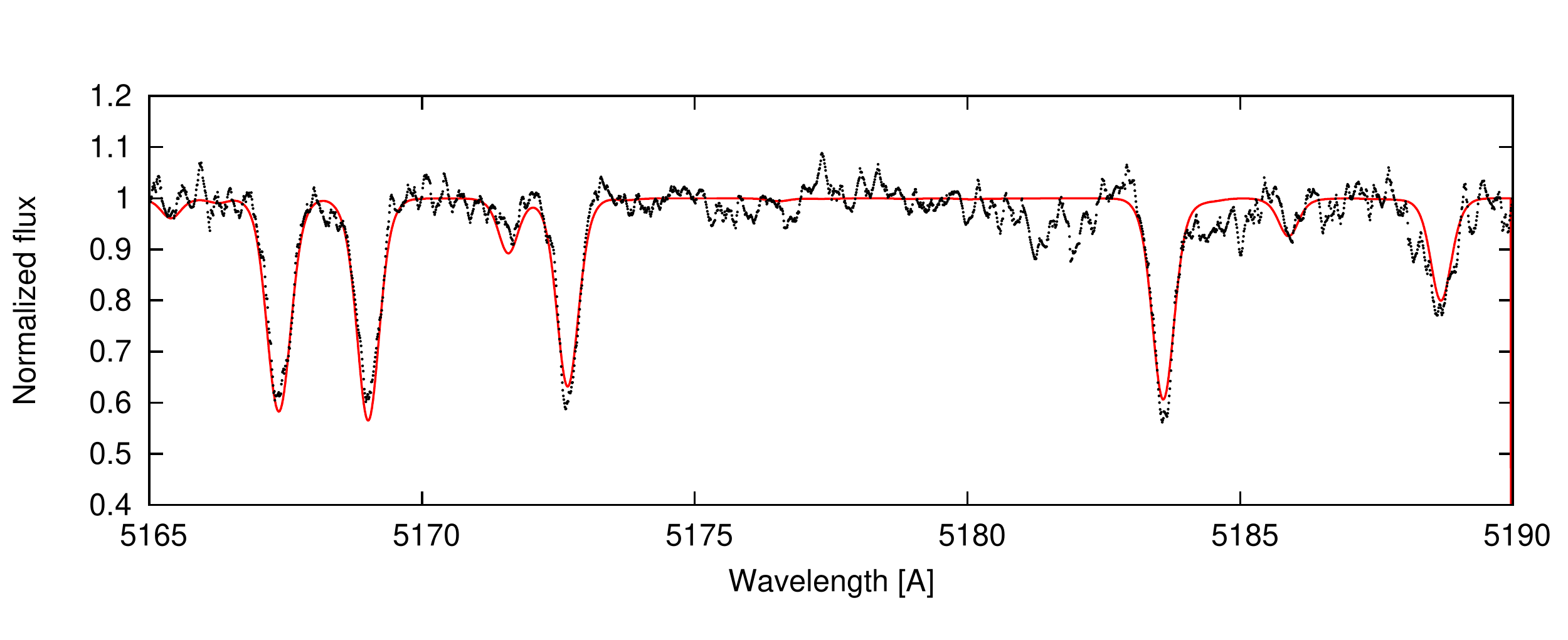}%
\caption{
Spectral fit for the Mg triplet range of T\,Sex.
The small black dots show the median spectrum over the whole pulsation cycle.
The red continuous line shows the best-fitting synthetic spectrum. 
}\label{fig:Mg_fit}%
\end{figure}

We also performed a more quantitative estimation for macroturbulent velocity 
by fitting theoretical stellar model atmospheres to the observed spectra.
Following \citet{Sneden17}, we prepared a good S/N combined spectrum from the best 16 spectra, 
which samples the complete pulsation cycle almost evenly. Then we shifted each of them 
with the corresponding radial velocity
and computed the median spectrum.
The vicinity of the Mg triplet (between 5165 and 5190~{\text\AA}) of the median spectrum is 
shown in Fig.~\ref{fig:Mg_fit} with black dots.
We fitted theoretical spectra between 5150 and 5190~{\text\AA} to this median spectrum by minimizing $\chi^2$.

We calculated the atmospheric parameters of T\,Sex from this region 
using the `Synthetic spectral fitting' tool of {\sc{iSpec}}.
After trying several radiative transfer codes (SPECTRUM, Turbospectrum, MOOG) integrated into the 
{\sc{iSpec}} package, we concluded that there is no significant difference in the obtained fits in our case,
so the integrated {\sc{SPECTRUM}} radiative transfer code \citep{Gray94}
and MARCS GES model atmospheres \citep{Gustafsson08} were selected.
The solar abundance of \citet{Grevesse07} and GES (Gaia-ESO Survey) atomic line list \citep{Heiter15} were used.

Several test runs showed that the parameters of \citet{Sneden17} obtained from the phase-averaged spectrum agrees within the errors with the results from our median spectrum,
therefore, we accepted and updated them. We accepted $T_{\mathrm{eff}}=6960 \pm 160$~K  effective temperature, 
$\log g=2.12 \pm 0.16$, [M/Fe]\,=\,$-1.48\pm0.1$ metallicity and [$\alpha$/Fe]\,=\,$0.51\pm0.18$ alpha-element enhancement.
However, we use our higher value for the microturbulence ($\xi=3.8$~km~s$^{-1}$)
because it always resulted in better spectral fits than $\xi=2.3\pm 0.16$~km~s$^{-1}$ of \citet{Sneden17}. Note that the errors correspond to one $\sigma$ uncertainties.

The macroturbulence $v_{\mathrm {mac}}$ and the projected rotational velocity $v\sin i$ lead to similar line broadening effects, 
as \citet{Preston19} showed, and
 it is difficult to separate them. We arrived at the same conclusion here. 
A similarly accurate fit can be achieved either with  
$v\sin i = 0$, $v_{\mathrm {mac}} = 23$~km~s$^{-1}$ or with $v\sin i = 15$, 
$v_{\mathrm{mac}} = 0$~km~s$^{-1}$, respectively. 
Since RR~Lyrae stars are slow rotators, it is very likely that the first case is closer to reality. This fit is shown by a red line 
in Fig.~\ref{fig:Mg_fit}.

Although in a completely different way, we arrived at the same conclusion as before by studying the variations in the H$\alpha$ line:
either $v\sin i$ or rather $v_{\mathrm {mac}}$ is relatively 
large ($\sim$15-20~km~s$^{-1}$). 
This supports our phenomenological explanation of the periodic line distortions.

\subsection{Sodium lines}

\subsubsection{Line doubling}

\begin{figure*}%
    \centering
    \includegraphics[width=8.5cm, trim=3cm 0cm 3cm 0]{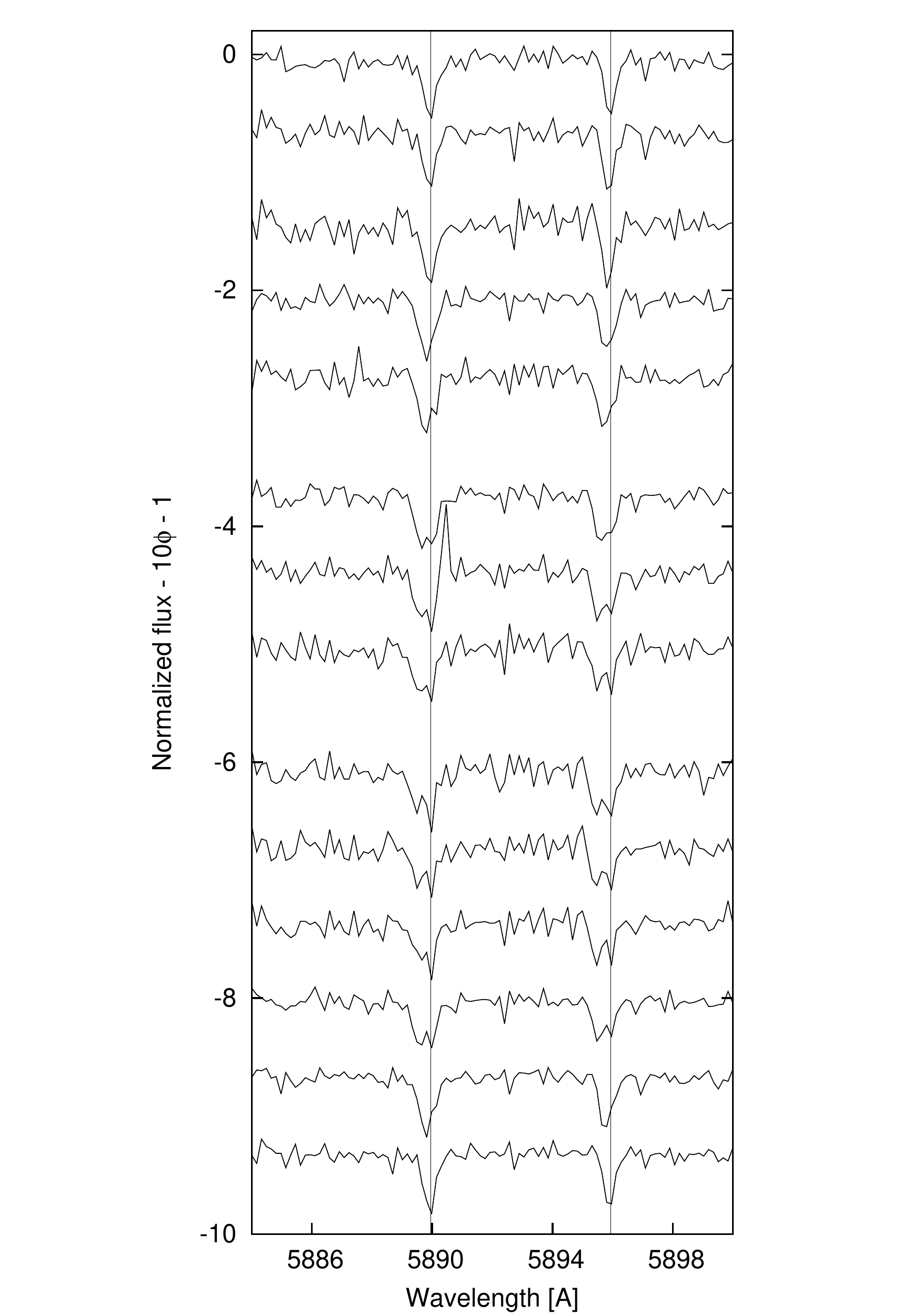}%
    \includegraphics[width=8.5cm, trim=3cm 0cm 3cm 0]{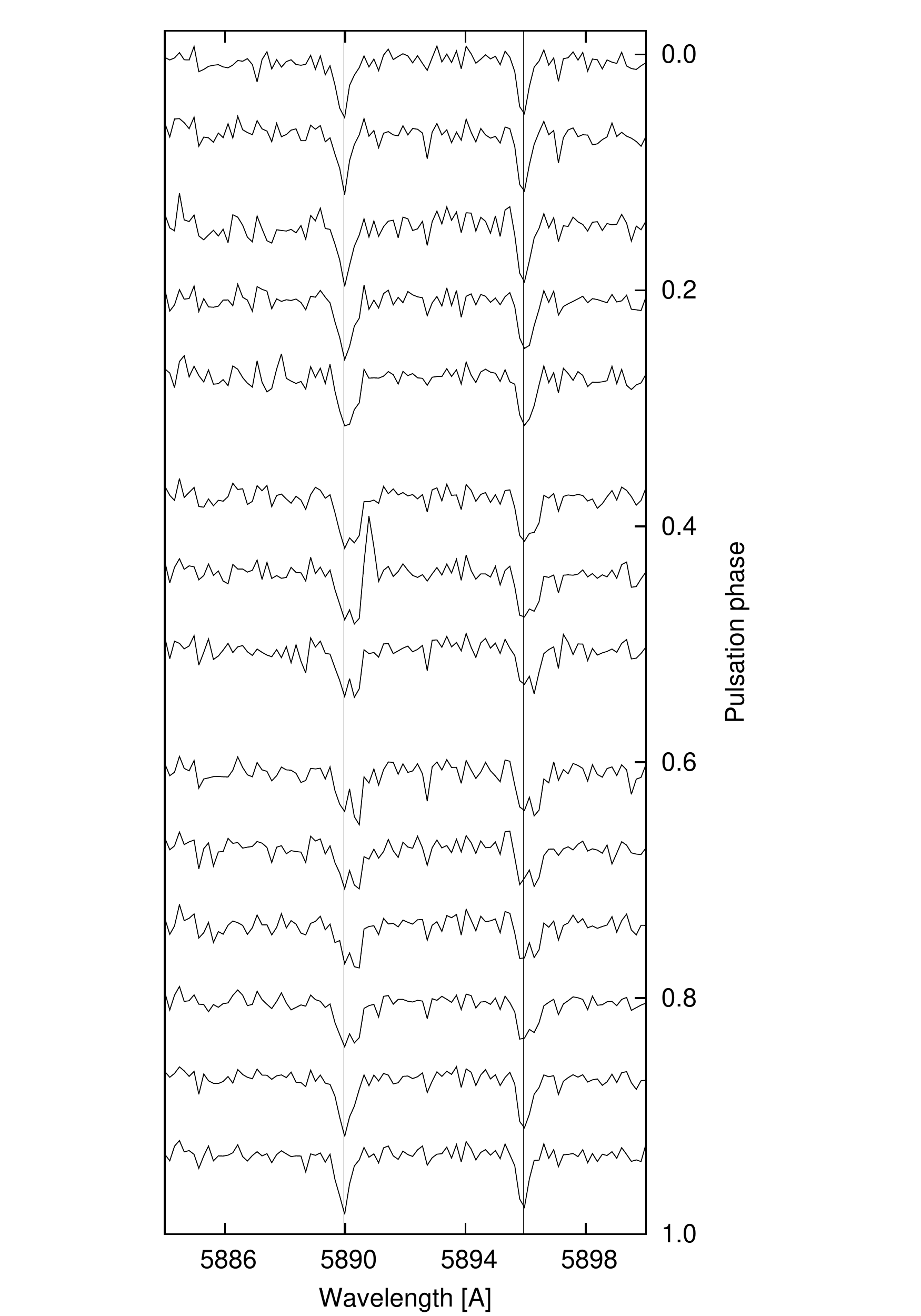}
\caption{Sodium line variations of T\,Sex over the pulsation cycle in a system co-moving with the pulsating atmosphere (in the left)
and in the stellar centre-of-mass rest frame (in the right), respectively.
The smoothed and normalised spectra are shifted vertically according to the pulsation phase. 
The thin vertical lines indicate the laboratory wavelengths of the Na D$_1$ and D$_2$ lines. Spectra in the right-hand panel are shifted with the radial velocity of the stellar centre-of-mass ($-13.2$~km~s$^{-1}$) 
to transform the Na~D lines to the vicinity of their laboratory wavelengths.
The slightly different line shapes are due to differences in the spline interpolation.}\label{fig:Tsex_NaD}%
\end{figure*}

We show the Na~I D line-profile variations of T\,Sex over the pulsation cycle in Fig.~\ref{fig:Tsex_NaD}.
In the left panel the spectra are plotted in a system co-moving with the pulsation, that is, the instantaneous radial velocity was subtracted from each spectrum.
In other words, this system moves with the average pulsation motion of the atmosphere
defined by the radial velocity curve in Fig.~\ref{fig:vrad}.
It can be seen that the D$_1$ and D$_2$ line profiles are rather similar in all phases:
they show significant deviation from a single Gaussian profile
between the phases of $\phi \approx 0.28$ and $\phi \approx 0.88$. The distortion
becomes line doubling between $\phi \approx 0.37$ and $\phi \approx 0.80$.
The phenomenon is observable over 60\% of the pulsation cycle.  

Line doubling of the sodium D lines in the spectrum of a fundamental-mode pulsating RRab star
(RR\,Lyr itself) was reported for the first time
by  \citet{Gillet17}, who found this line doubling to be coincident with an H$\alpha$
emission of RR\,Lyr at the phase $\phi \approx 0.227$. In a more recent and 
detailed study, \citet{Gillet19} showed that the D$_1$ line is doubled over 
75\% of the pulsation cycle. The position of the redder component is fixed during the whole pulsation cycle within the stellar rest frame,
therefore this component was explained by interstellar origin.  

Our present study shows line doubling phenomenon in an overtone pulsating RRc star.
As we see in the left-hand panel of Fig.~\ref{fig:Tsex_NaD},
the position of the red part of the line is fixed with respect to the co-moving frame, that is, 
it follows the atmospheric motions, which are represented
by the radial velocity curve and obtained from the averaged motions of metallic lines.
Thus, it can be assumed that the atmospheric layer where this line is formed is at or near
the layer where the metallic lines taken into account in the calculation of the radial velocity 
curve are formed.

The position of the blue component in the left panel of Fig.~\ref{fig:Tsex_NaD} varies with the phase 
but this is a virtual variation. When we construct the spectral variations over the
pulsation cycle within a rest of frame of the center-of-mass of T~Sex (right panel of Fig.~\ref{fig:Tsex_NaD}),
we see that actually the position of the blue component is fixed. In other words this component does not share the 
motion of the pulsating atmosphere. 

What is the origin of these `fixed' components? The telluric origin is unlikely because the
known telluric absorption lines in this spectral region (see e.g. \citealt{Hobbs78}) 
are weak and they have complex fine structure which has not been detected.
Circumstellar and interstellar origin are the two natural potential explanations.
According to \citet{Gillet19}, the $+$50.3~km/s velocity interstellar line they found 
in the spectra of RR\,Lyr  may originate 
from the wall of the Local Bubble \citep{Frisch11}. This explanation applies for our case as well.
The recent 3D-map of the interstellar gas
based on Na (and Ca) absorption observations \citep{Vergely10,Welsh10} shows a
gas cloud toward the direction of T\,Sex ($l=235^{\circ}38', b=+40^{\circ}36'$).
This cloud is also part of the material that forms of the wall of the Local Bubble.
Although the map includes only the 300~pc neighbourhood of
the Sun, the contribution of more distant matter to the
interstellar absorption is likely to be small, 
since T\,Sex have a high Galactic latitude, where we do not expect significant
interstellar matter so far away.

Distinguishing between circumstellar and interstellar material is generally an
observing task. We have to observe several stars that are 
close to the target star in space, and if we observe similar interstellar absorption
in the check stars, we can conclude that these lines originate from the interstellar space.
Otherwise, if only the target shows these lines, that suggests a circumstellar origin.
The method was successfully applied for discovering circumstellar discs 
(see e.g. \citealt{Redfield07a, Redfield07b, Rebollido18}). In the case of distant sources beyond the Local Bubble, the
picture could be more complicated. As \citet{Points04} showed, each selected star of the $\chi$ and h Per double cluster show different 
Na~I D absorption properties. This reflects the fine structure of the interstellar material in the line of sight.
The distance of T\,Sex ($803\pm40$~pc,  
from the Gaia DR2 \citealt{Gaia16, Gaia18}) is much larger than the
wall of the Local Bubble ($\sim 100$\,--\,200~pc), but, 
due to its high galactic latitude ($l=+40^{\circ}$), no significant interstellar material is expected beyond the Local Bubble.
\begin{figure}%
    \centering
    \includegraphics[width=3.5cm,angle=270,trim=0cm 0cm 0cm 0cm]{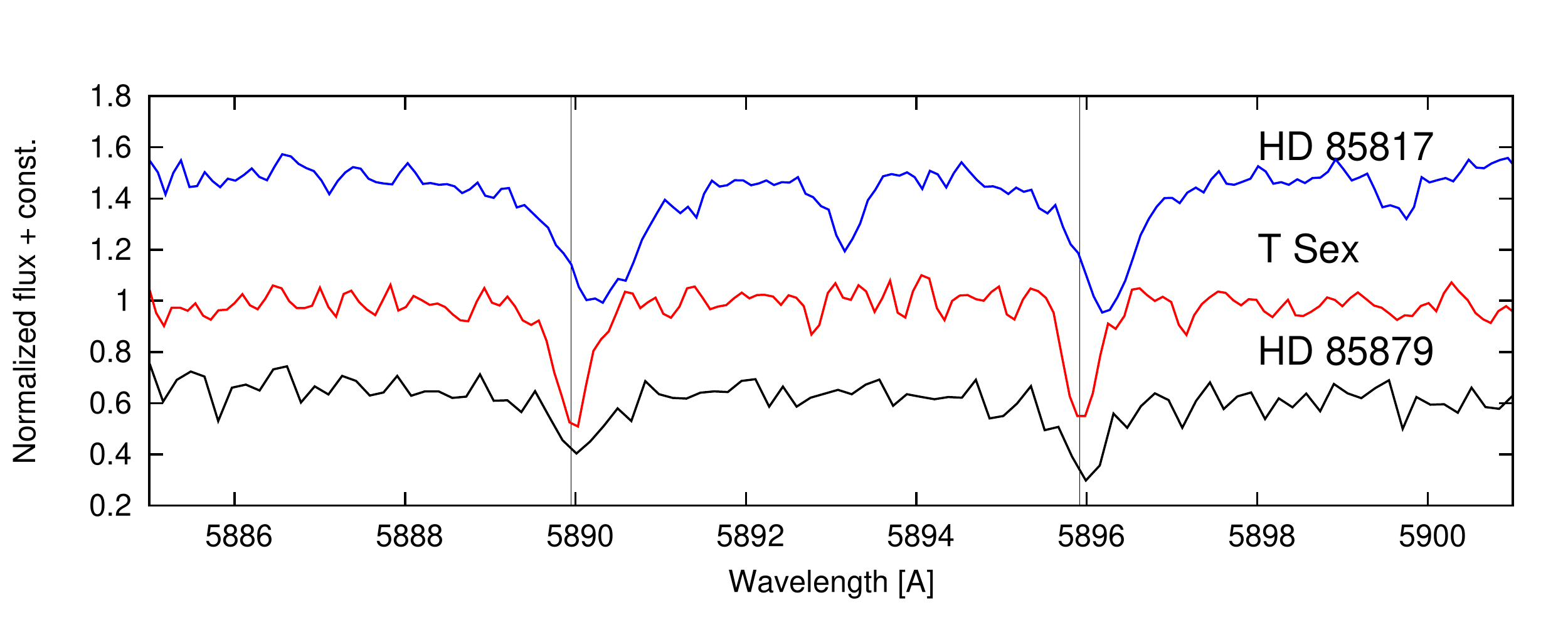}%
\caption{Spectra around the Na~D lines of T\,Sex and two neighbouring stars; HD\,85879 and HD\,85817. In each case, five individual spectra
were combined and equal shifts (with $18.5$~kms$^{-1}$) were applied to transform the wavelengths of the Na lines to the laboratory values.
}\label{fig:interstellar}%
\end{figure}

In the February of 2020, we observed 5\,--\,5 spectra of two brighter stars (HD\,85879 and HD\,85817) appearing close to the celestial position of T\,Sex, to investigate the issue. We determined their radial velocities with the same method as we 
described for T\,Sex in Sect.~\ref{sec:vrad}.
The obtained values are $16.01\pm0.34$~kms$^{-1}$ and $29.19\pm0.37$~kms$^{-1}$ for HD\,85879 and HD\,85817, respectively. Both values are in agreement
with the Gaia results \citep{Gaia18}.

We selected five spectra for T\,Sex from those phases where there is no line doubling (between $\phi=0.87$ and 0.14), i.e. the fixed components of the Na lines coincide with the moving components. 
The shift value $31.7-13.2=18.5$~kms$^{-1}$ obtained from Fig.~\ref{fig:Tsex_NaD} meaning the barycentric velocity of
the (interstellar) component was also applied for all three spectra. 
After averaging the 4-4 best S/N spectra for all three stars, the result is plotted in Fig.~\ref{fig:interstellar}.

The wide line profiles of HD\,85817 (top blue curve) do not show line doubling, only on the red wing of both lines of Na~D doublet 
slight breaks can be suspected just in the proper position. These asymmetric profiles might be the result of a blend between 
the star and interstellar lines.    
Unfortunately, the radial velocity difference between HD\,85879 and the interstellar matter is small ($\sim 2.5$~kms$^{-1}$), therefore, the lines coming from the star and from the interstellar material can not be distinguished (bottom black curve).
Although these control measurements did not provide a definite answer to the question, they rather support, in particular those of HD\,85817, that we are dealing here with material of interstellar and not circumstellar origin.

\subsubsection{P Cyg profile?}

At the phase $\phi=0.438$ T\,Sex shows an 
emission peak redwards to the red component of Na~D$_2$ line (see Fig.~\ref{fig:Tsex_NaD})
and it forms a P Cyg profile with the absorption line. 
As we have seen above this redder line component comes from the
stellar photosphere. 
Appearance of sodium emission is remarkable because such phenomenon has not
been reported for any RR\,Lyrae star before.

We investigated possible sources of this spectral feature.
The first possibility to exclude was an observational or reduction error. 
This emission line profile is constructed by 6 data points 
which means that the spectral resolution is much higher than the line width. 
Therefore, it is highly unlikely that the line would be an incorrectly treated cosmics.
Perhaps a more spacious anomaly in the CCD frame could cause this. However,
as we checked it, no such discrepancies (e.g. hot column, scattered light, 
saturation trail) are seen in the raw observed frame.
Additionally, there is no extra flux in any of the orders below and above the order
which shows the line.

It is possible that the a given emission line does not come from the stars, but somewhere
from the Earth's atmosphere.  The peak intensity of our 
$\lambda5890$ line is $\sim$72 per-cents
higher than the strongest known optical telluric emission line of O~I at $\lambda5577$.
The intensity of telluric sodium D$_2$ line is much weaker 5 per-cents of   
this atmospheric O~I emission line \citep{Louistisserand87}.
Furthermore, in the case of terrestrial origin, the phenomenon should be detected 
more or less continuously. 
These arguments make instrumental or terrestrial origin very unlikely.

Emission of a metallic line for RR\,Lyr stars has been completely unknown so far.
H$\alpha$ emission is known for three phase intervals of RRab stars. 
The H$\alpha$ emission just before the luminosity maximum ($\phi \sim 0.89-0.93$)
was detected long ago \citep{Struve48, Stanford49}. Later,
\citet{Gillet88} found a blue emission shoulder of the H$\alpha$ line.
This emission appears around $\phi \sim 0.73$, which is the position of the `bump' near the
photometric minimum. The third emission effect, 
a red emission shoulder around $\phi \sim 0.18 - 0.39$ was discovered by \citet{Preston11}.

Beyond the hydrogen emission phases of RRab stars,
helium emissions were also discovered by \citet{Preston09} in a sample of 11 observed
RRab stars. A stronger emission of He~I D$_3$ line at $\lambda5875.66$ and a weaker emission
at $\lambda6678.16$ were observed in all sample stars during rising light.
A further helium emission of He~II at $\lambda4685.68$ was also reported by \citet{Preston11}.

For all of these cases, the physical explanation of the emission is similar: different shock waves in 
the atmosphere (see \citealt{GilletFokin14} and references therein).
But similarly to the observations, the theoretical efforts have also been concentrated on
fundamental-mode pulsating RRab stars. We were not able to find any theoretical study
that investigates shock properties in RRc stars. 
Maybe the reason is that because of the smaller atmospheric velocities of RRc stars,
no strong shocks are expected. From an observational point of view, \citet{Sneden17} found
no signs of shocks (line doubling, emission) in the spectra of 7 RRc stars, and estimated
an upper limit of $\sim10$~km~s$^{-1}$ for any possible atmospheric shocks.  
At the same time, the authors discussed the possibility of a compression wave 
around $\phi\sim0.52\pm0.05$ where certain H$\alpha$ radial-velocity curves 
show a secondary maximum. As they mentioned, this wave would be ``qualitatively similar to the shock 
Sh$_{\mathrm {PM3}}$ of RRab stars \citep{Chadid14}, which produces compression heating during infall."

If we look into Gautschy's latest theoretical calculations (see fig.~4 in \citealt{Gautschy19})
phase $\phi\sim0.4-0.5$\footnote{
We mention that Gautschy denotes the maximum radius by phase $\varphi=0.0=1.0$ and not 
the maximum brightness as we use here. There is a $\sim0.68$ phase shift between the two zero-points.} 
of RRc stars corresponds to the phase where the early shock appears in the RRab stars \citep{Hill72}.
Namely, where the luminosity functions of the cool edge and the hot edge
of the He~II partially ionised zones intersect.
The physical meaning of this intersection is the collision between the infalling high atmosphere 
and the slower shrinking photosphere. 

Based on these, our finding of Na emission in the appropriate pulsation phase both 
agrees with the theoretical calculations and the preliminary observational expectations.
The wavelength difference between the absorption and emission components of D$_2$ line
is  0.46~\text{\AA}, giving a velocity difference of 23.4~km~s$^{-1}$.
Sodium absorption lines, along with other metallic lines, are formed practically
in the photosphere. The velocity of the photosphere at this phase is 5.8~km~s$^{-1}$.
The total velocity of the supposed compression wave in the stellar rest frame is 29.2~km~s$^{-1}$.
This is almost an order of magnitude smaller than the shock wave velocities estimated for RRab stars.

Our detection of this emission rises two questions.
First, why did not we observe similar emission at line D$_1$?
And second, why did not \citet{Sneden17} detect it in their very good time-sampled spectra? 
There might be an answer to the first question: the intensity ratio between Na emission  
D$_1$ and D$_2$ lines can be strongly varying between 0.5\,--\,0.9 depending on the physical condition of the emitting material
(e.g. \citealt{Nikaidou83, Slanger05}). The explanation of the second question may be that 
the phenomenon is temporary. This would not be unprecedented, since the cycle-to-cycle 
variation of RR Lyrae spectra is a known phenomenon \citep{Chadid00}. 
For more definite answers, we need more spectroscopic time series observations.

\section{Summary}

In this paper, we presented our results 
based on new photometric and spectroscopic time series
on one of the brightest RRc star, T\,Sex.
Our main findings are:
\begin{itemize}
\item{
We found from the Fourier analysis of the photometric data set of the {\it TESS} space telescope that the light curve of T\,Sex can be described by two independent 
frequencies: the frequency of the radial overtone pulsation, $f_1$, and a frequency
belonging most probably to an $l=8$ non-radial pulsation mode. The speciality of the Fourier
spectrum of T\,Sex is that the usual $f_x$ frequency is not significant, only $0.5f_x$ is observed.
No similar stars have been reported before by space photometry.
Such stars are also very rare in ground-based observations (seven stars among 960, $\sim 0.7\%$, \citealt{Netzel19}).
}

\item{
Based on our time-resolved spectroscopic measurements, we showed a 
characteristic phase-dependent periodic distortion of the H$\alpha$ line.
This type of line-profile variation is significantly different from 
RRab stars, e.g., from the almost unchanging profiles seen at RR\,Lyr \citep{Gillet19}.
This type of line-profile changes have not previously been published for RR\,Lyrae stars, 
only for other radial pulsators \citep{Nardetto08,Nardetto13}.
The phenomenon is most likely caused by the relatively high macroturbulent 
velocity of T~Sex ($\sim 15$\,--\,$20$~km~s$^{-1}$), which can be caused by convection. 
This plays a more important role in the pulsation of RRc stars than in RRab pulsators.
}

\item{We discovered a phase-dependent line doubling of the Na~D lines.
This is the first case that 
such line doubling has been reported for overtone pulsating RR\,Lyrae stars. 
The possible explanation of the feature is the same as in the case of RR\,Lyr \citep{Gillet19}:
one of the line
components come from the star, while the other one originates from the interstellar medium.
This finding calls the attention to the fact that the characteristic velocities of 
the atmosphere of an RR\,Lyrae star during pulsation is similar to the velocity of the ISM clouds 
in the wall of the Local Bubble. Thus, it is expected that most Galactic field RR Lyr stars 
will show similar Na~D line doubling.
}

\item{
At the phase $\phi=0.438$, 
a definite emission peak was found on the red side of the sodium  D$_2$ line, 
which forms a P\,Cyg profile with the absorption component. 
The appearance of P\,Cyg profiles in RR Lyrae spectra
is usually associated with shock waves. 
Although no strong shock waves are expected for RRc stars, the appropriate phase of 
the detected event and the calculated shock wave velocity ($\sim 30$~km~s$^{-1}$) 
also suggest the appearance of a weak shock wave. Due to the single detection, 
the intrinsic nature of the phenomenon must be verified in the future.
}

\end{itemize}

\section*{Acknowledgements}

This work was supported by the Hungarian National Research, 
Development and Innovation Office by the Grant NN-129075
and the Lend\"ulet Program of the Hungarian Academy of Sciences, 
project No. LP2018-7/2018. JMB thanks to Dr A. Mo\'or and Dr R. Szab\'o for their valuable suggestions.

This work has made use of data from the European Space Agency (ESA) mission
{\it Gaia} (\url{https://www.cosmos.esa.int/gaia}), processed by the {\it Gaia}
Data Processing and Analysis Consortium (DPAC,
\url{https://www.cosmos.esa.int/web/gaia/dpac/consortium}). Funding for the DPAC
has been provided by national institutions, in particular the institutions
participating in the {\it Gaia} Multilateral Agreement.

\section*{Data availability}

The raw and processed spectroscopic data underlying this article will be
shared on reasonable request to the corresponding author.
The radial velocity data obtained from the spectra are available in the
online supplementary material under the name of {\tt tsex\_vrad.dat}.
The photometric data underlying this article were accessed either from ASAS-3 database \url{http://www.astrouw.edu.pl/asas/?page=aasc&catsrc=asas3} and from Mikulski Archive for Space Telescopes \url{https://mast.stsci.edu/portal/Mashup/Clients/Mast/Portal.html} (TESS). The derived data generated in this research will be shared on reasonable request to the corresponding author.






\bsp	
\label{lastpage}

\end{document}